\documentstyle[preprint,prl,aps,floats]{revtex}
\input{psfig.tex}

\tighten
\begin{document} 
\draft
\twocolumn[\hsize\textwidth\columnwidth\hsize\csname@twocolumnfalse\endcsname
\preprint{Imperial/TP/96-97/51}
\title{The case against scaling defect models of cosmic structure formation}
\author{Andreas Albrecht, Richard A. Battye and James Robinson}
\address{Theoretical Physics Group, Blackett Laboratory, Imperial
College, Prince Consort Road, 
 London SW7 2BZ,  U.K.\\}
\maketitle
\begin{abstract}
We calculate predictions from defect models of
structure formation for both the matter and Cosmic Microwave
Background (CMB) over all observable scales.  Our results point to  
a serious problem reconciling the observed large-scale galaxy
distribution with the COBE 
normalization, a result which is robust  
for a wide range of defect parameters. 
We conclude that standard scaling defect models
are in conflict with the data, and show how attempts to
resolve the problem by considering non-scaling defects would require
radical departures from the standard scaling picture. 
\end{abstract}

\date{\today}

]

\renewcommand{\thefootnote}{\arabic{footnote}}
\setcounter{footnote}{0}

Defect models offer an elegant explanation of the origin of cosmic
structure.  The idea is that some distribution of defects---or more
generally, field disorder---is produced during a  cosmic phase transition\cite{kib}.  The defects then start a process of
`coarsening' which continues through the present day, 
contributing a component to the matter in the universe which evolves
in a highly non-linear way.  Cosmic strings, for example, move at
relativistic velocities, periodically self-intersect  and break off
loops, 
which themselves eventually decay into gravity waves. Such processes
can seed the onset of gravitational collapse in a universe which is
initially perfectly homogeneous.  

In contrast with other models of cosmic structure
formation, calculations for defects require the modelling of
highly non-linear processes from very
early times (e.g. the time of Grand Unification),
right up to the present day. During this  period the 
universe increases by around 25 orders of magnitude in size.
 Only recently  has it become practical to solve
the full Boltzmann equations for the matter and radiation
perturbations in the presence of defect sources consistently modelled
over such a length of time\cite{PSelTa}. 
Accurate large-scale numerical simulations are currently
the best source of information
on the details of the defect evolution, but 
even using state of the art technology it is still necessary to
extrapolate with scaling arguments 
\cite{kib} to achieve anything like the required dynamic
range.  Here we describe work that is 
not as closely linked to specific defect simulations and is therefore
able to explore a wider range of possible defect models. Thus we can
systematically investigate the robustness of the clash between defect
models and observations. 

Our calculations use the fact that if only the 
power spectra are to be calculated, then one only needs two-point
functions of the defect stress energy as input\cite{AS}. Scaling
arguments can then be used to increase the dynamic range. This 
scaling behavior has been observed to some degree in numerical
simulations\cite{AT,BB,AShe,PST,DZ} and has become part of the
standard lore of defect evolution, although the extent to which
it is valid over a factor of $10^{25}$ in cosmic expansion is not
yet clear. 

For the current calculations this approach was incorporated into a
version of CMBFAST\cite{cmbfast} modified to include source stress
energy for the scalar, tensor and vector contributions, which are
generic in defect based models\cite{PSelTa,PSelTb,ACDKSS}. In order to
do this, we model the components of the defect stress energy under a
number of simple assumptions which maintain causality and conserve
stress energy. The source is approximated by a network of line-like
segments with correlation length $\xi\eta$ at conformal time $\eta$ and
velocity taken from a gaussian distribution with RMS $v$, truncated to
prevent $v>c$. The number of lines is reduced causally, so as to
maintain a constant density with respect to the horizon.  This approach is
similar to the model used in ref.\cite{VHS}, which was shown to give
two point functions in good agreement with string simulations for
certain stress energy components, but we 
have updated it to include all the components required and an improved
decay mechanism\cite{ABRb}.  

Any active source which creates perturbations inside the horizon is
likely to be 
incoherent, leading to the absence or suppression of secondary Doppler
peaks\cite{ACFM}. Hence, the form of unequal-time correlators (UETC)
are also important. Our approach, which contrasts with that used in
ref.\cite{PSelTa}, is not to calculate the UETC directly, but to
create an ensemble of source histories with the correct two-point
correlation statistics. Then in order to calculate the ensemble
average of the matter power spectrum and the CMB, one must use the
Boltzmann code for each source history and average the resulting
spectra. The results shown here used 200-400 realizations which gives very
small statistical errors, but runs with just 40 realizations sufficed
to establish the basic picture. 

\begin{figure}[t]
\centerline{\psfig{file=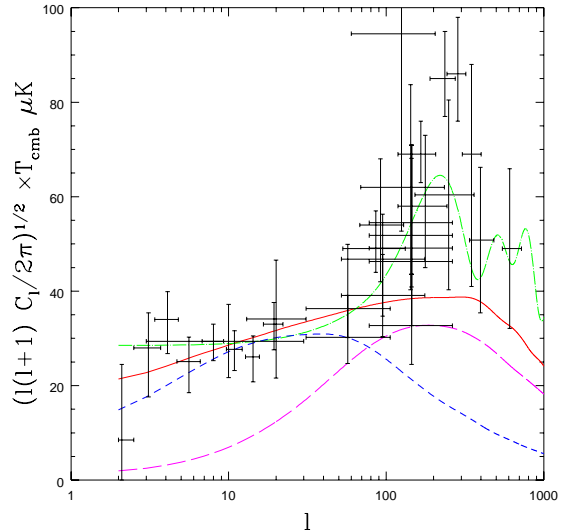,width=2.95in}}
\caption{The (COBE normalized) angular power spectrum of CMB
anisotropies for the standard cosmic string model (solid) plotted with
the current observational data, the standard CDM curve (dotted). The
two dashed curves give the partial contributions from two time windows
to either side of $z=100$} 
\label{cmb1}
\end{figure}

In Fig.~\ref{cmb1} we plot the angular power spectrum of the CMB for
what we shall call the standard string model (solid line). This uses
string model parameters $\xi=0.3$ and $v=0.65$ as suggested by
simulations,  an assumption of perfect scaling from defect formation
to the present day and a flat background cosmology with $\Omega_{\rm 
c}=0.95$, $\Omega_{\rm b}=0.05$ and $h=0.5$ where $H_0=100h\,{\rm km
}\,{\rm sec}^{-1} 
{\rm Mpc}^{-1}$. Included also are the standard CDM model  based on
inflation (dot-dashed line) and all the current published data-points 
with error-bars based on the assumption of
gaussianity\cite{tegmark}. The main 
features to note are the absence of any discernible Doppler peak in
the defect spectrum and the apparent conflict with the data points.
We have repeated 
these calculations for various different values of $\xi$ and $v$ and
also for sensible variations of the cosmological parameters $h$ and
$\Omega_{\rm b}$. The spectrum is modified by these variations, but
none manage  to increase the amount of power at angular scales with
l=200-400 by very much. Clearly the situation looks bad for defect
models, although it is worth noting that the plotted error bars are
one-sigma, and deviations from the assumed gaussianity may require
even larger error bars due to the small sky coverage.   We expect the
situation to be much clearer when the new CMB data arrives in the near future.

Fig.~\ref{cmb1} also shows the partial
results which come from integrating the defect contributions over two time
windows: Window 1
(z = 1300 to z = 100, during which $\eta$ increases by a factor of 5)
gives the long-dashed curve and window 
2 (z = 100 to z = 1.6, during which $\eta$ increases by a factor of 7)
 gives the short-dashed curve.  This information
will be helpful in the subsequent discussion of non-scaling defect models.

\begin{figure}[t]
\centerline{\psfig{file=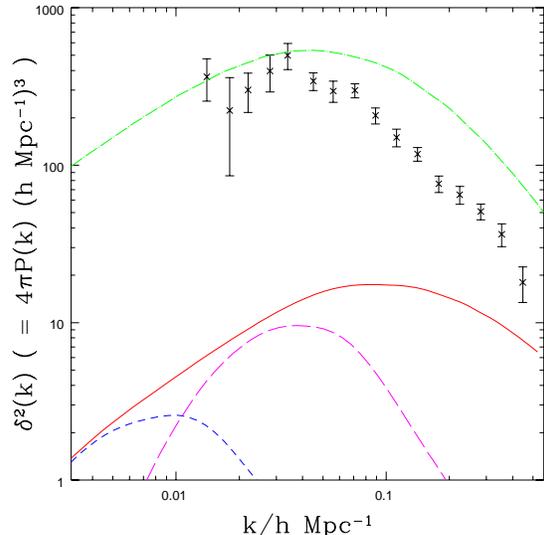,width=2.95in}}
\caption{The power spectrum of the dark matter perturbations for the
same models and windows shown in Fig. 1, plotted with the data.} 
\label{mat1}
\end{figure}

\begin{figure}[t]
\centerline{\psfig{file=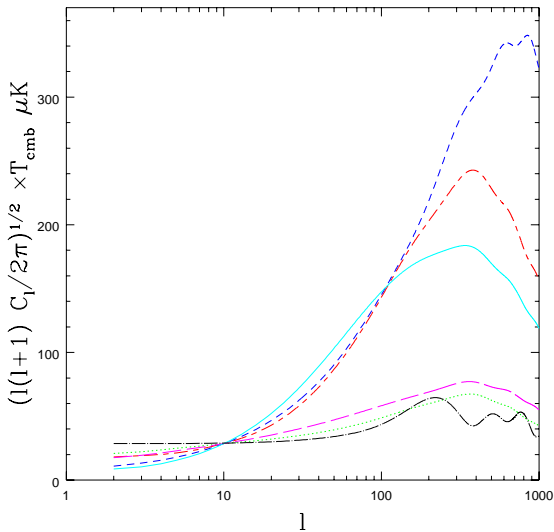,width=2.95in}}
\caption{The angular power spectrum of CMB anisotropies for the
various non-scaling models discussed in the text.  The three most
extreme models (which have reasonable values of $b_{100}$) have the
highest peaks.  Standard CDM is included for reference  (dash-dot
curve).}    
\label{cmb2}
\end{figure}

\begin{figure}[t]
\centerline{\psfig{file=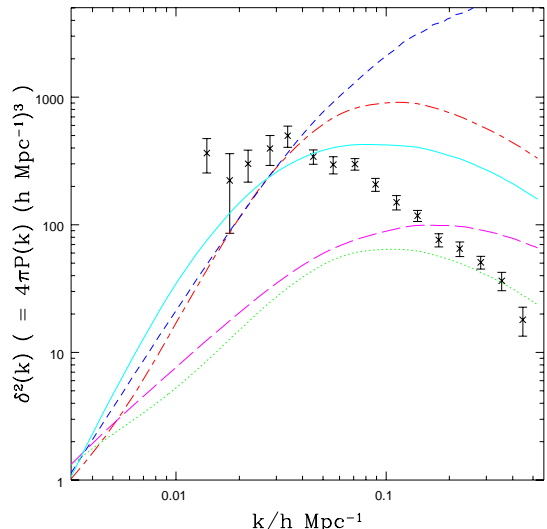,width=2.95in}}
\caption{The power spectrum of the dark matter perturbations plotted with
the data for the same models as Fig.~\ref{cmb2}} 
\label{mat2}
\end{figure}

Using the same annotation as Fig.~\ref{cmb1}, Fig.~\ref{mat1} shows
the COBE normalized cold dark matter density perturbation power 
spectrum predicted for the standard string model and that
for  standard CDM, compared with the data\cite{PD}. The contributions
from the same two time windows are 
also included, as in Fig.~1.  Theory and data are often compared using
$\sigma_8$, the variance of the fractional matter overdensity
in a ball of radius $8h^{-1}{\rm Mpc}$. For standard CDM
$\sigma_8=1.2$ for $h=0.5$, while the value favoured by observations is
$\sigma_8=0.5$. If one were to compare with  the string
model at these 
scales, one calculates $\sigma_8=0.26$ and hence the bias on
these scales between the galaxy distribution, which is largely
baryonic matter, and the Cold dark matter is $b_8=\sigma_8/\sigma_8^{\rm DM}
\approx 2$. This not unreasonable value can even
be slightly reduced by changing the string model and
cosmological parameters. 

However, these comparisons ignore the fact that there is a woeful
absence of power on  larger 
scales. We quantify the conflict for low $k$ by calculating  the
hypothetical bias $b_{100} 
\equiv \sigma_{100}/\sigma_{100}^{\rm DM} $ where $\sigma_{100}$ is
defined in analogy to $\sigma_8$, but for spheres of $100h^{-1}{\rm
Mpc}$ and the favoured value $(\sigma_{100}=3.5\times 10^{-2})$ is  
calculated for a smooth curve which gives a good fit to the data
points.  The standard string model has 
$b_{100}= 5.4$ which cannot be improved substantially by any of the
variations
already discussed.  The chances of a real physical model having such a 
large value of $b_{100}$ are remote\cite{Wandelt}, and there is no
observational evidence for a large $b_{100}$\cite{GF,WSDK}. We
conclude that the standard string model is 
in conflict with the observations at an unacceptable level on scales
around $100h^{-1}{\rm Mpc}$. Variations relating to
possible systematic uncertainties in simulations and our knowledge of
the baryon density and Hubble constant cannot alleviate the
discrepancy.  
We also note that these conclusions do not depend on the
stringy nature of the model that we have used. For instance, the
results are very similar
if we impose a sharp sub-horizon cut-off on the
source stress energy, mimicking behaviour closer to that of cosmic
textures.

By far the most effective way of addressing the large $b_{100}$ problem is to
exploit the uncertainty in the overall scaling behavior of the
string network.  It is this behavior, 
after all, that relates the contributions from defects on different
scales, and has simply been put into our calculation by
hand.  Although scaling has been observed to some degree in
simulations, it is not completely clear that simple laws are
valid over a wide dynamic range and events during the history of the
universe may lead to deviations. For example, the
radiation-matter transition is known at the very least to 
cause a shift; one could speculate that
this is not yet well understood.  

We have extensively probed the possible deviations from the standard
picture, and found it very difficult to get around the large 
$b_{100}$ problem.  This can be understood by looking at the
contributions from the two time windows illustrated in
Figs.~\ref{cmb1} and \ref{mat1}. The first window provides
essentially 
all the contributions to the COBE normalization, while the second
window provides the dominant contribution to $\sigma_{100}$.  The problem is
that these two windows span a sufficiently narrow period in the defect
history that  
something dramatic must happen to the scaling behavior to shift their
relative contributions sufficiently.   An extreme (and un-motivated)
`solution' which suggests itself is to simply turn off the string
network at $z=100$, hence preserving perturbations which contribute to
$\sigma_{100}$, but removing the highest possible fraction of
the contributions to COBE scales.
The result of doing this is illustrated in
Figs.~\ref{cmb2} and \ref{mat2} (solid line), and manages to give
$b_{100} = 1.2$. 

There are two simple types of deviations from scaling which may be more
acceptable. The first, which has been observed to occur to
some degree at the
radiation-matter 
transition\cite{MS}, is just a step in the string density
from one value to another, occurring in a smooth way over some period
of time. Such a deviation can be quantified by the ratio
$\chi=(\eta^{1/2}\theta_{00})^{\rm{rad}}/(\eta^{1/2}\theta_{00})^{\rm{mat}}$ 
(where $\chi =1$ gives ``standard scaling"). For 
instance, if the radiation-matter transition gives rise to a
difference in  the amount of small scale structure on the strings 
in the two eras, then $\chi$ could be interpreted as the ratio of
the renormalized 
string tensions. The
second type of
deviation  we consider 
is a power law deviation from scaling quantified by a parameter
$\alpha$ via $\theta_{00}\sim
\eta^{-(\frac{1}{2}+\alpha)}$, for which the density in strings
$\rho\sim 1/\eta^{2+2\alpha}$, with the choice $\alpha =0$ corresponding
to a standard scaling law. This may model the behaviour, for example,
in an open universe or in one 
dominated by a cosmological constant \cite{M,VB}.  

In order to illustrate the problem, we show the
results from four further models for the CMB in Fig.~3 and for the
matter power spectrum in Fig.~4. The first two are mild deviations
from scaling which one might imagine are plausible: Model A (dotted
curve, $b_{100} = 3.4$) is a transition of $\chi=2$, with the
transition beginning at $8\eta_{\rm eq}$ and ending at
$10\eta_{\rm eq}$ where $\eta=\eta_{\rm eq}$ is the time of equal
matter-radiation. Model 
B  (long-dash curve, $b_{100} = 2.9$) is a power law deviation from
scaling with $\alpha=0.25$. 

The other two examples are much more extreme; their virtue
being that they can fit the data points in the matter power spectrum
at around $100h^{-1}{\rm Mpc}$: Model C (long-short dashed curve,
$b_{100} = 1.0$) is a 
transition between the same times as for model A but with $\chi=10$ and  
Model D (short-dashed curve, $b_{100} = 0.7$) is a power law deviation from
scaling with $\alpha=0.75$. 
Whilst models C and D fit the matter spectrum at scales of around
$100h^{-1}{\rm Mpc}$, they completely fail to fit smaller scale galaxy
data, and 
give a large excess of small scale power in the CMB
spectrum. One might hope that changes to the ionization history and
matter content of the universe could solve some of these problems. However,
the fact remains that there is no evidence to suggest that such
extreme deviations from scaling could occur in the standard defect scenario.

We now relate our results to previous work on the subject.   Several
papers have discussed the bias in COBE normalized defect models.
In ref.\cite{PST} a serious bias problem was noted on scales up to $20
h^{-1}$ Mpc, but concerns remained that the simulations were not
including  all the relevant contributions (particularly to the density
fluctuations) because of their limited dynamic range.
The compilation of refs.\cite{AS} and \cite{ACSSV} in ref.\cite{WS}, although
looking very much like our Fig. 2, involved very different treatments
of the defects at scales relevant to COBE vs $b_{100}$, and it is not
clear that a straightforward compilation is valid. 
Our work avoids these uncertainties by solving the full Boltzmann
equations with a single source model to compute 
perturbations consistently on all scales.  The scaling assumption
translates into essentially `infinite' dynamic range. 
 
There has also been work more recently which is on equal footing in this 
respect\cite{PSelTa}.  Conceptually, the main difference between this
work and ours is that they extract the UETC's directly from numerical
defect simulations. Both groups scale the correlation functions to
gain dynamic range. In addition, ref. \cite{PSelTa} uses some sophisticated
methods to work efficiently with the UETCs. Bearing in mind these
differences, it should be noted that the 
two results look very similar. A strength of the simulation based
approach is that the UETC's are associated with well defined defect
scenarios. On the other hand, our approach is more flexible, allowing
us to explore a wide range of UETC's in order to test the robustness  
of the results against variation amongst defect models as
well as possible systematic uncertainties in the simulations.

Even at a more technical level there is a large degree of similarity
between our results and those in ref.\cite{PSelTa}.  We 
find that on super-horizon scales the scalar, vector and tensor
anisotropic stresses are in the simple ratio  
$\langle |\Theta^S|^2 \rangle : \langle |\Theta^V_i|^2 \rangle :
\langle |\Theta_{ij}^T|^2\rangle = 3:2:4 $
as imposed by causality and isotropy. Around $l=10$, 
we find that $C_l^{S}:C_l^{V}:C_l^{T}$ approximately $3 : 1 : 0.4$.  Our
value of $C_l^{V}:C_l^{T}$ is very close to that in ref.\cite{PSelTa},
while our $C_l^{S}:C_l^{V}$ is somewhat larger.  The
degree of similarity is striking given that we use a simple model
while large simulations were used in ref.\cite{PSelTa}.
We believe our larger $C_l^{S}:C_l^{V}$ is due to the
relatively large value of $\Theta_{00}$ 
compared to  $\Theta^{S}$ in our model.
We have not directly compared our sources with those in \cite{ACDKSS}
(describing local strings), but our scalar component
seems to be larger.  These differences (and those between
\cite{ACDKSS} and \cite{PSelTa}) suggest that the relative strength
of the scalar component may vary noticeably from one type of
defect to another. 
We
should note that even if there were only a scalar component the
comparison with the current 
data would be very bleak; the vector and tensor components only make
things worse.  While it has been suggested that models with
highly suppressed anisotropic stresses might achieve 
improved values for the COBE normalized bias\cite{DS},
no concrete defect model has been proposed which has this feature. 

Therefore, we conclude that standard scaling defect scenarios are ruled out on
the basis of current data, and that this situation can only be
remedied by extreme modifications to the scaling law.

We thank U. Seljak and M. Zaldariagga for the use of CMBFAST, and in
particular Seljak for help with 
incorporating the vectors and tensors. We thank
M. Hindmarsh, L. Knox, P. Shellard, P. Ferreira, G.
Vincent and  N. Turok for helpful conversations. This work was
supported by PPARC and computations
were done at the UK National Cosmology
Supercomputing Center, supported by PPARC, HEFCE and Silicon
Graphics/Cray Research. RAB is funded by PPARC grant GR/K94799.

\def\jnl#1#2#3#4#5#6{\hang{#1, {\it #4\/} {\bf #5}, #6 (#2).}}
\def\jnltwo#1#2#3#4#5#6#7#8{\hang{#1, {\it #4\/} {\bf #5}, #6; {\it
ibid} {\bf #7} #8 (#2).}} 
\def\prep#1#2#3#4{\hang{#1, #4.}} 
\def\proc#1#2#3#4#5#6{{#1 [#2], in {\it #4\/}, #5, eds.\ (#6).}}
\def\book#1#2#3#4{\hang{#1, {\it #3\/} (#4, #2).}}
\def\jnlerr#1#2#3#4#5#6#7#8{\hang{#1 [#2], {\it #4\/} {\bf #5}, #6.
{Erratum:} {\it #4\/} {\bf #7}, #8.}}
\def\prl{Phys.\ Rev.\ Lett.}
\def\pr{Phys.\ Rev.}
\def\pl{Phys.\ Lett.}
\def\np{Nucl.\ Phys.}
\def\prp{Phys.\ Rep.}
\def\rmp{Rev.\ Mod.\ Phys.}
\def\cmp{Comm.\ Math.\ Phys.}
\def\mpl{Mod.\ Phys.\ Lett.}
\def\apj{Ap.\ J.}
\def\apjl{Ap.\ J.\ Lett.}
\def\aap{Astron.\ Ap.}
\def\cqg{Class.\ Quant.\ Grav.} 
\def\grg{Gen.\ Rel.\ Grav.}
\def\mn{MNRAS}
\def\ptp{Prog.\ Theor.\ Phys.}
\def\jetp{Sov.\ Phys.\ JETP}
\def\jetpl{JETP Lett.}
\def\jmp{J.\ Math.\ Phys.}
\def\zpc{Z.\ Phys.\ C}
\def\cupress{Cambridge University Press}
\def\pup{Princeton University Press}
\def\wss{World Scientific, Singapore}
\def\oup{Oxford University Press}

\pagebreak
\pagestyle{empty}

\end{document}